\documentclass[a4paper,twocolumn]{esapub} % European paper
\usepackage[dvips]{graphicx}
\usepackage{natbib}
\bibpunct{(}{)}{;}{a}{}{;}
\usepackage{times}

\title{ Flarelike brightenings of active region loops observed with SUMER}
\author{T. J.  Wang, D. E. Innes,  S. K. Solanki, and W. Curdt }
\affil{Max-Planck-Institut f\"ur Sonnensystemforschung, 37191 Katlenburg-Lindau,
Germany\\
email:  wangtj@mps.mpg.de }

\begin{document}

\maketitle

\footnotemark
\footnotetext{Proceedings of the 11th European Solar Physics Meeting "The Dynamic Sun: Challenges for Theory and Observations" (ESA SP-600). 11-16 September 2005, Leuven, Belgium. Editors: D. Danesy, S. Poedts, A. De Groof and J. Andries. Published on CDROM., id.105.1}


\begin{abstract}
 Coronal loops on the east limb of the Sun were observed by SUMER on SOHO for
several days. Small flare-like brightenings are detected
very frequently in the hot flare line Fe~{\small XIX}. 
We find that the relatively intense events
are in good coincidence with the transient brightenings seen by Yohkoh/SXT. 
A statistical analysis shows that these brightenings have durations of 5-84 min
and extensions along the slit of 2-67 Mm. The integrated energy observed 
in Fe~{\small XIX} 
for each event is in the range of $3\times10^{18}-5\times10^{23}$ ergs, and 
the estimated thermal energy ranges from $10^{26}-10^{29}$ ergs. Application of the 
statistical method proposed by Parnell \& Jupp (2000) yields  a value of 1.5 to 1.8
for the index of a power law relation between the frequency of the events and the
radiated energy in Fe~{\small XIX}, and a value of 1.7 to 1.8 for the index of the
frequency distribution of the thermal energy in the energy range
$>10^{27}$ ergs. We examine the possibility that these small 
brightenings give a big contribution to heating of the active region corona.
\end{abstract}

\section{Introduction}
It is well known that there is not enough energy in large flares to heat the corona
because they are so rare that their time-averaged power contribution is too small.
However, a multitude of small-scale reconnection events (microflares or nanoflares)
may be a possible source for heating the corona \citep[e.g.,][]{lev74, par88}. 
Balloon-borne observations of many small hard X-ray bursts with energies between 
$10^{24}-10^{27}$ ergs by \citet{lin84}
seem to support this idea. Systematic detection and statistical analysis
of small-scale phenomena in the corona have been explored in recent years with
soft X-ray and EUV high-resolution imaging telescopes such as Yohkoh/SXT, SOHO/EIT 
and TRACE \citep[see a review by][]{asc04}. Active region transient brightenings 
(ARTBs) \citep[e.g.][]{shi92} and X-ray bright points (XBP) in the quiet Sun
\citep[e.g.][]{gol74} are governed by flare-like processes and 
release thermal energy comparable to hard X-ray microflares. EUV transient
brightenings in the quiet Sun with temperatures of 1-2 MK and energies of 
$10^{24}-10^{27}$ ergs were first studied with SOHO/EIT data \citep[e.g.][]{kru98} and 
with TRACE data \citep[e.g.][]{par00}. It appears that these transient EUV brightenings
have all the properties of larger flares, indicative of a continuous distribution
of energy release in the corona from large flares down to tiny nanoflares. 

The frequency distribution of the energy release 
can be used to examine the microflare or nanoflare
heating hypothesis. \citet{hud91} showed that in order to explain the heating
of the corona by microflares or nanoflares, the frequency distribution of small
energy release events must be a power law with index steeper than 2. In many
of the early papers \citep[see a review by][]{cro93}, the index of the frequency
of regular flares was estimated to be around 1.8 by assuming that the energy in
a flare is linearly related to the peak flux of the flare. \citet{cro93} calculated
the distribution of the total nonthermal energy using hard X-ray bremsstrahlung
observations, and found a value of 1.53 for the index of event energies in the
range $10^{28}-10^{31}$ ergs. \citet{shi95} studied ARTBs with energies in the
range $10^{27}-10^{29}$ ergs, and they estimated the index for the energy 
distribution to be 1.5-1.6. For the EUV nanoflares detected by SOHO/EIT and
TRACE, many estimates were made for the index with results between 1.4 and 2.6
\citep[see reviews by][]{asc04, par04}. This clearly shows 
nanoflare heating of the corona remains an open question.  

In this paper, we study hot flare-like brightenings which frequently occur in coronal 
loops as seen in SUMER spectra taken 10 to 40 Mm above active
regions on the limb. We first make a multi-wavelength analysis of these events, then
we make a statistical study of their physical properties and calculate frequency
distributions of their peak intensities and energies to explore whether the
contribution of these events is important in heating the active-region corona.

\begin{figure}
\centering
\includegraphics[width=0.8\linewidth]{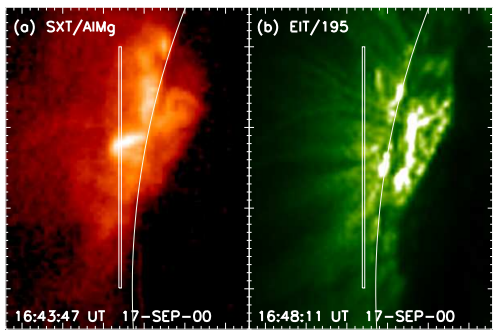}
\caption{\label{arimg}
 {\bf a)} Image of loop systems in AR~9167 and AR~9169 observed with SXT
AlMgMn-sandwich filter. The offlimb position of the SUMER slit is marked as a narrow box.
{\bf b)} The same region observed with EIT 195\AA~filter.}
\end{figure}

\section{Observations}
The data sets analyzed include two observations which were made by the SUMER 
spectrometer in the sit-and-stare mode (see Fig.~\ref{arimg}).
The first observation was recorded with a cadence of
90 s and the $300{''}\times4{''}$ slit during 16$-$20 September 2000. Five spectral lines
were transmitted, including Fe~{\small XIX} $\lambda$\,1118 (6.3 MK), 
Ca~{\small XV} $\lambda$\,1098 and $\lambda$\,555$\times$2 (3.5 MK), Ca~{\small XIII}
 $\lambda$\,1134 (2.2 MK), and Si~{\small III} $\lambda$\,1113 (0.06 MK). 
The second observation was made with a cadence of 162.5 s 
during 25$-$30 September 2000. A transmitted spectral window covering 1097$-$1119
\AA~contains a number of lines formed in the temperature range 0.01$-$10 MK, e.g.
Fe~{\small XIX} $\lambda$\,1118, Ca~{\small XV} $\lambda$\,1098, Al~{\small XI}
 $\lambda$\,550$\times$2 (1.6 MK), Ca~{\small X} $\lambda$\,557$\times$2 (0.7 MK), 
Ne~{\small VI} $\lambda$\,558$\times$2 (0.3 MK), and Si~{\small III} $\lambda$\,1113.

\section{Results: Multiwavelength analysis}
\begin{figure}
\centering
\includegraphics[width=0.8\linewidth]{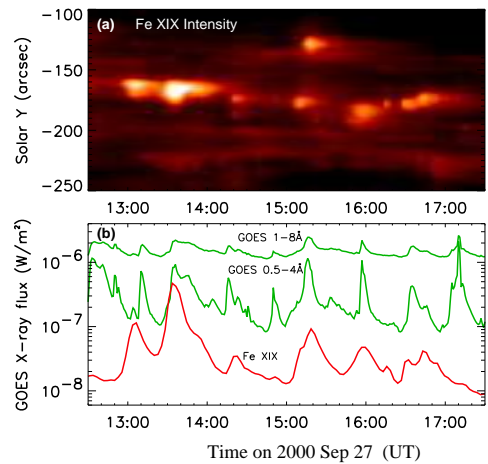}
\caption{\label{sumgs}
   {\bf a)} Intensity time series of the Fe~{\small XIX} line detected at the upper
part of the slit. {\bf b)} Light curve of the Fe~{\small XIX} line integrated 
along the slit in arbitrary units. Light curves of GOES full-sun soft X-ray flux 
through 1-8 \AA~and 0.5-4 \AA~are also plotted. The flux of GOES 0.5-4 \AA~is 
multiplied by a factor of 10.}
\end{figure}

\begin{figure}
\centering
\includegraphics[width=0.8\linewidth, height=0.7\linewidth]{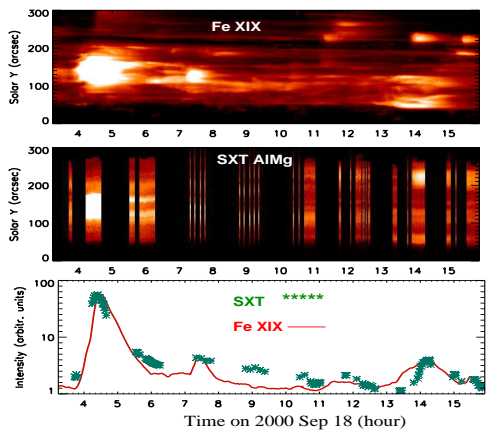}
\caption{\label{sumsx}
{\bf a)} Intensity time series of the Fe~{\small XIX} line detected at the slit.
{\bf b)} Same as {\bf a)} but for the SXT AlMgMn-sandwich filter.
{\bf c)} Light curves of the Fe~{\small XIX} line and the SXT AlMgMn filter
integrated over the slit.}
 \end{figure}

We first compare Fe~{\small XIX} Brightenings with GOES Flares. Figure~\ref{sumgs} 
shows such an example. We find that 40 of 53 Fe~{\small XIX} enhancements, which are 
identified from light curves on 16-19 Sep and 25-28 Sep 2000, coincide in time with 
GOES C-class flares. Note that many of these brightenings are associated with 
longitudinal loop oscillations seen in the Fe~{\small XIX} line \citep{inn04}.
We find that all major brightenings seen in Fe~{\small XIX} are also clearly seen 
in SXT emission, indicating that the emission seen by SUMER in the Fe~{\small XIX} 
line and the SXT brightenings are signatures of the same phenomena. 
Figure~\ref{sumsx} shows such an example. However, a long succession of SUMER
observations is preferable to do a complete statistical study, over SXT data with
gaps.

\begin{figure}
\centering
\includegraphics[width=0.8\linewidth, height=0.8\linewidth]{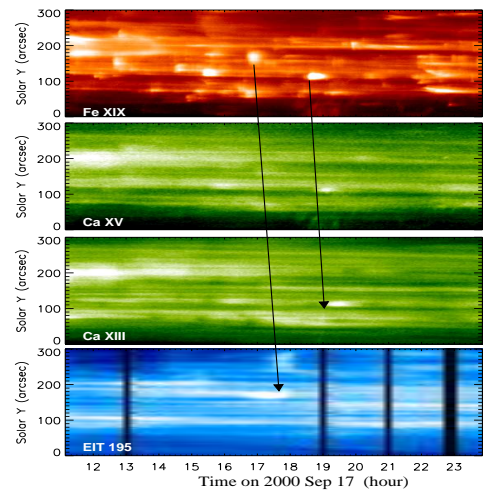}
\caption{\label{cool}
  Line-integrated intensity time series of {\bf a)} the Fe~{\small XIX} line, {\bf b)}
the Ca~{\small XV} line, {\bf c)} the Ca~{\small XIII} line, and {\bf d)}
the EIT 195 \AA~filter, detected
at the slit. The arrows demonstrate the cooling of two hot brightenings seen
in Fe~{\small XIX} and with a time delay as low-temperatures brightenings.}
\end{figure}

Figure~\ref{cool} illustrates the cooling of Fe~{\small XIX} brightenings 
into the 195 \AA~EIT filter (1.5 MK). For 12 strong Fe {\small XIX} brightenings, 
Their average cooling times from Fe~{\small XIX} to Ca~{\small XV}, 
Ca~{\small XIII} (or Al~{\small XI}), and EIT 195\AA~filter temperatures 
are 27 min, 53 min, and 70 min for limb loops at a projected height of 
$\sim$30 Mm.  We find that most of the Fe~{\small XIX}
brightenings have no counterparts seen in the low ($<$4 MK) temperature lines.
There are two explanations. One is due to the obscuration by strong
background emission in the normal coronal lines. The other is because 
coronal loops in the AR core are generally at 5-8 MK \citep{yos96},
so do not cool to lower temperatures.

\section{Results: Statistical analysis}

\subsection{Identification of Brightening Events by Automatic Search}

We first remove the background emission and determine the noise level ($\sigma$) 
at each pixel position. In order to make an automatic search, we define 
the following criterion for an event:\\
1) $I_{dif}>N\sigma$, where $I_{dif}$ is the enhancement of peak intensity relative to
the minimum intensities preceding and following the event. This condition defines
an event for a single pixel.\\
2) $\Delta{T_p}\leq N\Delta{t}$, where $\Delta{T_p}$ is the time difference between  
peak intensities at two neighboring pixels, $\Delta{t}$ is the time cadence. This 
step groups together the neighboring spatial pixels which peak almost simultaneously.\\ 
3) $L_y\geq~N$ pixels, where $L_y$ is the extension of events along the slit. In this
step, we separate the neighboring events along the slit according to their peak intensity
variation, and define the smallest events which should fulfil the given condition.

\subsection{Physical Condition of Brightenings}
\begin{figure}
\centering
\includegraphics[width=0.8\linewidth, height=0.3\linewidth]{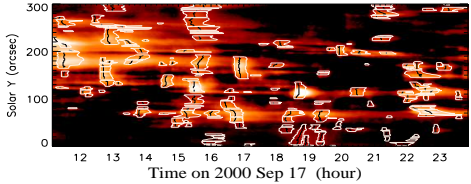}
\caption{\label{atsch}
Example of the Fe~{\small XIX} brightening identification with automatic search.
The black curves mark the peak
times of intensity in an event. The time series of the Fe~{\small XIX} intensity shown here
is the same as that in Fig.~\ref{cool}, but with the background removed.}
\end{figure}

We have identified in total 1334 events matching the criteria $I_{dif}>1\sigma$, 
$\Delta{T_p}\leq 3\Delta{t}$, and $L_y\geq~3$ pixels from the two data sets  
with a total observing duration of 96.6 hours.
Figure~\ref{atsch} shows such an example. We calculate the duration of an event at 
a single pixel as the FWHM of an enhancement in intensity relative to its preceding 
minimum. We define the duration of each event as the average value of durations for 
all spatial pixels in this event and define the extension of each event along the slit 
as the scale of all grouped pixels making this event. We obtain a mean duration 
of 22$\pm$13 min ranging from 5-84 min, and a mean extension $L_y$ of
7$\pm$6 Mm ranging from 2-67 Mm.
The peak intensity of events has a mean of 0.03$\pm$0.12 W/m$^2$
in the range 10$^{-4}$-2.2 W/m$^2$. Under the assumption that the temperature at peak
intensity is at a value of 6.3 MK, i.e. the temperature where the emissivity
reaches the maximum ($G_{max}$), we can estimate the emission measure by
$I_{peak}/G_{max}$. We calculated that the emission measure of brightenings has a mean
of (0.3$\pm1.4)\times10^{28}$ cm$^{-5}$ in the range 
10$^{25}$-2.5$\times10^{29}$ cm$^{-5}$.
Assuming a constant line-of-sight depth of 10 Mm, we estimated that the electron density
has a mean of (1.1$\pm1.3)\times10^{9}$ cm$^{-3}$ in the range 10$^8$-1.6$\times10^{10}$
cm$^{-3}$. To estimate the thermal energy content for each event, we need to know the loop
geometry. Assuming that the temperature of brightening loops is 6~MK, the loop length
($L$) is 3 times $L_y$ and the loop width ($w$) is 10 Mm, we estimate the thermal
energy by $E_{th}=3n_ek_BT(Lw^2)$. We obtained that $E_{th}$ has a mean of
(7.9$\pm19)\times10^{27}$ ergs in the range 1.8$\times10^{26}$-$2.5\times10^{29}$ ergs.
The integrated energy of brightenings observed in the Fe~{\small XIX} line has a mean of
(5.4$\pm27)\times10^{21}$ ergs in the range 2.8$\times10^{18}$-$5.1\times10^{23}$ ergs.

\subsection{Frequency Distributions}
Assuming that the distribution of the peak intensities or energies follows a power-law,
i.e.,
\begin{equation}
   f(E)=\frac{f_0}{E_0}\left(\frac{E}{E_0}\right)^{-\alpha},
\end{equation}
we determine the index, $\alpha$, by using the maximum likelihood method proposed by 
\citet{par00}. We obtained $\alpha\approx1.7$ for the peak intensity distribution
and  $\alpha$=1.5-1.8 for the frequency distribution of the radiated energy observed 
in Fe~{\small XIX}. For the thermal energies estimated based on the assumption of $L/L_y=3$,
we obtained $\alpha$=1.7-1.8 in the energy range greater than 10$^{27}$ ergs. 
We also made a test to calculate the thermal energy
of events by assuming $L/L_y=f_i$, where $f_i$ ($i$=1,2, ..., N) are taken from a
uniformly-distributed random series in the range 2-7. For 5 random series, we found
the values of $\alpha$ in the same range as the case of $L/L_y=3$. 

\begin{figure}
\centering
\includegraphics[width=0.8\linewidth, height=0.87\linewidth]{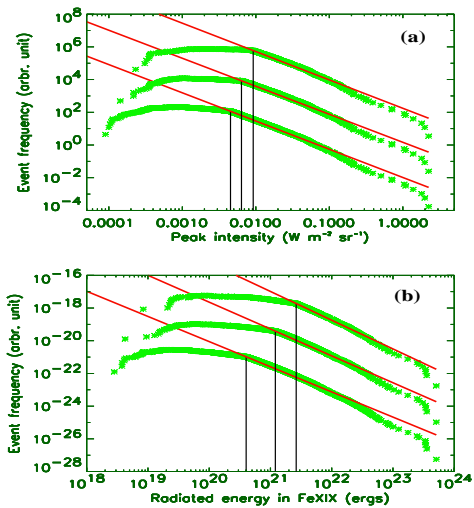}
\caption{\label{raderg}
Frequency of events vs. {\bf a)} event peak intensity, and {\bf b)} event
integrated energy in the Fe~{\small XIX} line for events of intensity enhancements at least
1$\sigma$, 2$\sigma$, and 3$\sigma$. These plots show the observed data (in green color) 
and the right-hand power law of the fitted skew-Laplace distribution (red line).
The frequencies of events for the three cases in {\bf a)} and {\bf b)} are multiplied
by 1, 10$^2$, 10$^4$, respectively, so they can all be drawn without overlap on the
same graph.}
\end{figure}

We can estimate the total energy rate by,
\begin{equation}
P=\int_{E_{min}}^{E_{max}} f(E)EdE. 
\end{equation}
If the distribution $f(E)$ is a power-law (see Eq.(1)), then we have
\begin{equation}
P=\frac{f_0E_0}{2-\alpha}\left(\left(\frac{E_{max}}{E_0}\right)^{2-\alpha}-\left(\frac{E_{min}}{E_0}\right)^{2-\alpha}\right).
\end{equation}
\noindent
Taking $\alpha=1.77$ for the case of $I_{dif}>1\sigma$ and $1.8\times10^{26}<E_{th}
<2.5\times10^{29}$ ergs for the observed events, we obtained $P=5.6\times10^{25}$ 
ergs s$^{-1}$. If assuming that the power-law continues from 10$^{24}$ to 10$^{33}$ ergs,
the upper limit of total energy supplied by these brightenings and flares was
estimated to be about 5$\times10^{26}$ ergs s$^{-1}$, or 5$\times10^{6}$ ergs 
cm$^{-2}$ s$^{-1}$ for ARs of the size of 100 Mm. This energy input rate is close to
but insufficient for the heating of the active-region corona, which has a typical
energy loss rate of 10$^7$ ergs cm$^{-2}$ s$^{-1}$ \citep{wit77}.

\section{Conclusion}
There are many heating events in active region loops that can only be seen at 
temperatures greater than 6 MK, e.g. microflares recently detected by RHESSI revealing
temperatures of 6-14 MK from the spectral fits in the 3-15 KeV energy range
\citep{ben02}. We see them clearly in SUMER Fe~{\small XIX} line emission. We estimate a
power-law index of 1.7-1.8 for the frequency distribution of the thermal energies
of these events in the range 10$^{27}$-10$^{29}$ ergs. Their energy input rate does
not seem to be sufficient for heating the active-region corona. 

\begin{figure}
\centering
\includegraphics[width=0.8\linewidth, height=0.45\linewidth]{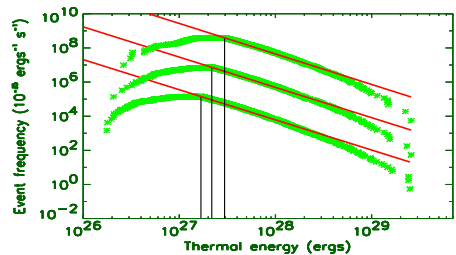}
\caption{\label{thmerg}
Frequency of events vs. event thermal energy. The thermal energy
of an event is calculated by assuming the loop length to be three times the
brightening extension along the slit. The frequencies of events for the three cases
(with intensity enhancements at least 1$\sigma$, 2$\sigma$, and 3$\sigma$) are multiplied
by 1, 10$^2$, 10$^4$, respectively.}
\end{figure}

\end{document}